\documentclass[11pt,a4paper]{article}
\pdfoutput=1
\usepackage{jheppub}
\usepackage[utf8]{inputenc}
\usepackage[english]{babel}
\makeatletter

\newcommand{\Tr}{\text{Tr}}

\title{Gravity from entanglement for boundary subregions}

\author[a]{David Blanco}
\author[a]{Mauricio Leston}
\author[b]{Guillem P\'erez-Nadal}

\affiliation[a]{\it Instituto de Astronom\'ia y F\'isica del Espacio, Universidad de Buenos Aires,\\
1428 Buenos Aires, Argentina}
\affiliation[b]{\it Departamento de F\'isica, FCEN, Universidad de Buenos Aires,\\
1428 Buenos Aires, Argentina}

\emailAdd{dblanco@iafe.uba.ar, mauricio@iafe.uba.ar, guillem@df.uba.ar}

\abstract{We explore several aspects of the relation between gravity and entanglement in the context of AdS/CFT, in the simple setting of 3 bulk dimensions. Specifically, we consider small perturbations of the AdS metric and the CFT vacuum state and study what can be learnt about the metric perturbation from the Ryu-Takayanagi (RT) formula alone. 
It is well-known that, if the RT formula holds for all boundary spacelike segments, then the metric perturbation satisfies the linearized Einstein equations throughout the bulk. 
We generalize this result by showing that, if the RT formula holds for all spacelike segments contained in a certain boundary region, then the metric perturbation satisfies the linearized Einstein equations in a corresponding bulk region (in fact, it is completely determined in that region).
We also argue that the same is true for small perturbations of the planar BTZ black hole and the CFT thermal state. We discuss the relation between our results and the ideas of subregion-subregion duality, and we point out that our argument also serves as a holographic proof of the linearized RT formula for boundary segments.}

\begin{document}

\maketitle

\section{Introduction}

According to the AdS/CFT conjecture \cite{Maldacena:1997re}, quantum gravity around Anti de Sitter (AdS) space is equivalent to a conformal field theory (CFT) on its boundary, 
in such a way that to each state of one theory corresponds a state of the other theory.
In the limit where the CFT is strongly coupled and has a large number of degrees of freedom  the gravity theory reduces to classical Einstein gravity, so in this limit different CFT states correspond to different classical spacetimes (supplemented with some configuration of the matter fields). A major goal in this context is to understand precisely how the bulk geometry (as well as other bulk physics) is encoded in the CFT state.

Significant advance in this direction has been triggered by the Ryu-Takayanagi (RT) formula for the entanglement entropies of the boundary CFT. In quantum field theory, the entanglement entropy of a spatial region $V$ is the von Neumann entropy of the reduced density matrix $\rho_V$ obtained by tracing out the degrees of freedom in the complement of $V$,
\begin{equation}
S_V=-\textrm{tr} \left(\rho_V \log \rho_V\right).
\label{entro}
\end{equation}
Note that this quantity depends on the state of the theory, and measures how this state entangles the region $V$ with its complement.
The RT proposal \cite{Ryu:2006bv}, in its generalized form due to \cite{Hubeny:2007xt}, asserts that, for theories with an Einstein gravity dual, the entanglement entropy is obtained by a simple geometric calculation,
\begin{equation}
S_V=\frac{1}{4G}\,\underset{v\sim V}{\textrm{ext}}\left[A(v)\right],\label{RTintro}
\end{equation}
where one extremizes the area $A(v)$ of the bulk surfaces $v$ that are homologous to the region $V$ in the boundary (if several extremal surfaces exist one picks the one with minimal area), and $G$ denotes Newton's constant. This proposal has passed numerous consistency checks \cite{Headrick:2013zda,Blanco:2013joa,Wall:2012uf}, and in fact it has been explicitly derived from holography
\cite{Casini:2011kv,Lewkowycz:2013nqa,Dong:2016hjy}.

Beyond its power as a tool for computing entanglement entropies (which are otherwise difficult to calculate, even in free field theories \cite{Casini:2009sr}), the RT formula provides deep insight into the workings of AdS/CFT. 
Indeed, given a CFT state one can view this formula as a constraint on the dual bulk geometry. This is a very strong constraint, so it is reasonable to expect that it may determine much, if not all, of the bulk geometry without any other input from holography. Thus, the RT formula suggests that much of the bulk geometry is encoded in the entanglement structure of the CFT state. Some explicit evidence for this idea was found in \cite{Swingle:2009bg,VanRaamsdonk:2009ar,VanRaamsdonk:2010pw}, where it was argued that boundary entanglement is responsible for the connectedness of the bulk spacetime. 

More evidence comes from the recent result that, at the linearized level, boundary entanglement is also responsible for bulk spacetime dynamics. Indeed, given a perturbation of the AdS geometry and the CFT vacuum state, the assumption that the RT formula holds to first order for all boundary spatial balls yields a set of nonlocal constraints on the metric perturbation which are exactly equivalent to the linearized Einstein equations \cite{Lashkari:2013koa, Faulkner:2013ica} (see also \cite{VanRaamsdonk:2016exw, Czech:2016tqr}). In fact, the analysis of \cite{Faulkner:2013ica} is more general: in it, the linearized equations of a generic (not necessarily Einstein) gravity theory are obtained from the corresponding holographic entanglement entropy formula, which is a Wald-like generalization of (\ref{RTintro}). These generic theories of gravity, which involve higher powers of the curvature, are dual to CFTs slightly away from the strong coupling limit; a similar generalization, which corresponds to moving slightly away from the large $N$ limit, was obtained in \cite{Swingle:2014uza}. The above result has been extended to second order in the metric and state perturbations in \cite{Faulkner:2017tkh}.

The RT formula has also played an important role in another, related development, namely the proposal that the way in which the bulk geometry and other bulk physics are encoded in the CFT state is ``local'', in the sense that a boundary domain of dependence contains complete information about some corresponding bulk region. There is a fair amount of evidence supporting this idea, usually referred to as subregion-subregion duality \cite{Bousso:2012sj,Bousso:2012mh,Czech:2012bh}, and it is currently believed that the bulk region associated to a given boundary domain of dependence $U$ is the so-called entanglement wedge of $U$ \cite{Czech:2012bh,Headrick:2014cta}. In this sense, the entanglement wedge reconstruction \cite{Dong:2016eik} gives an explicit example in which bulk operators can be reconstructed as CFT operators on a boundary domain of dependence, provided that they lie in the entanglement wedge. This is an improvement on other reconstruction methods which apply only to a smaller region in the bulk, known as the causal wedge \cite{Hamilton:2006az}.

The purpose of this paper is to obtain further evidence for the above ideas in a unified way, in the simple setting of 3 bulk dimensions. In the previously mentioned derivation of the linearized Einstein equations \cite{Lashkari:2013koa, Faulkner:2013ica,VanRaamsdonk:2016exw}, the RT formula is assumed to hold to first order for {\emph{all}} boundary spatial balls (which reduce to segments in the case where the boundary is 2-dimensional), and this results in the validity of the linearized Einstein equations throughout the bulk. With the ideas of subregion-subregion duality in mind, it is natural to ask if a ``local'' version of this result holds, i.e., if the linearized RT formula for spacelike segments contained in a certain boundary region implies the linearized Einstein equations in some corresponding bulk region. We will show that this is indeed the case. 

We will take the boundary region $U$ to be the domain of dependence of a spacelike segment, and we will see that the corresponding bulk region, where the linearized Einstein equations are satisfied, is the causal wedge $W(U)$, which in this case coincides with the entanglement wedge. In fact, we will show that the linearized RT formula for spacelike segments contained in $U$ not only implies the linearized Einstein equations in $W(U)$, but is equivalent to those equations supplemented with a boundary condition, which relates the metric perturbation to the state perturbation via the usual holographic formula for the expectation value of the CFT stress tensor. We will argue that this boundary value problem has a unique solution, thus providing an explicit example (at the linearized level) of bulk geometry emerging from boundary entanglement in a ``local'' way. We will also point out that the validity of the linearized RT formula for boundary segments in holographic theories follows from the above equivalence, so our arguments (which do not make use of the replica trick) also serve as an alternative holographic proof of that formula.
These results will be obtained not only for perturbations of the zero-temperature background (the background considered in \cite{Lashkari:2013koa, Faulkner:2013ica,VanRaamsdonk:2016exw}), but also at non-zero temperature.

\textbf{Outline}     \,\,\, The remainder of this paper is organized as follows.
In section \ref{ch:gravent} we prove the equivalence announced above for perturbations of the zero-temperature background. We do it in three steps: in subsection \ref{ch:segments} we find a pair of local equations which is equivalent to the condition that the linearized RT formula hold for all segments contained in some boundary spacelike segment; this result is then used in subsection \ref{ch:open} to show that the linearized RT formula for spacelike segments contained in a generic boundary open region $U$ is equivalent to the linearized Einstein equations in a corresponding bulk region $G(U)$ plus a standard holographic formula, which plays the role of a boundary condition; finally, in subsection \ref{ch:domains} we show that, in the case where $U$ is the domain of dependence of a spacelike segment, $G(U)=W(U)$, which completes the argument. In section \ref{ch:btz} we show that our zero-temperature results remain true for perturbations of a thermal background, and we close in section \ref{ch:discussion} with a discussion of our results.

\textbf{Note added:} as this work was nearing completion \cite{Lewkowycz:2018sgn} appeared, which presents some overlap with our results regarding the holographic derivation of the linearized RT formula without using the replica trick.

\section{Zero temperature}
\label{ch:gravent}

\begin{figure}[h]
\begin{center}
\includegraphics[scale=0.5]{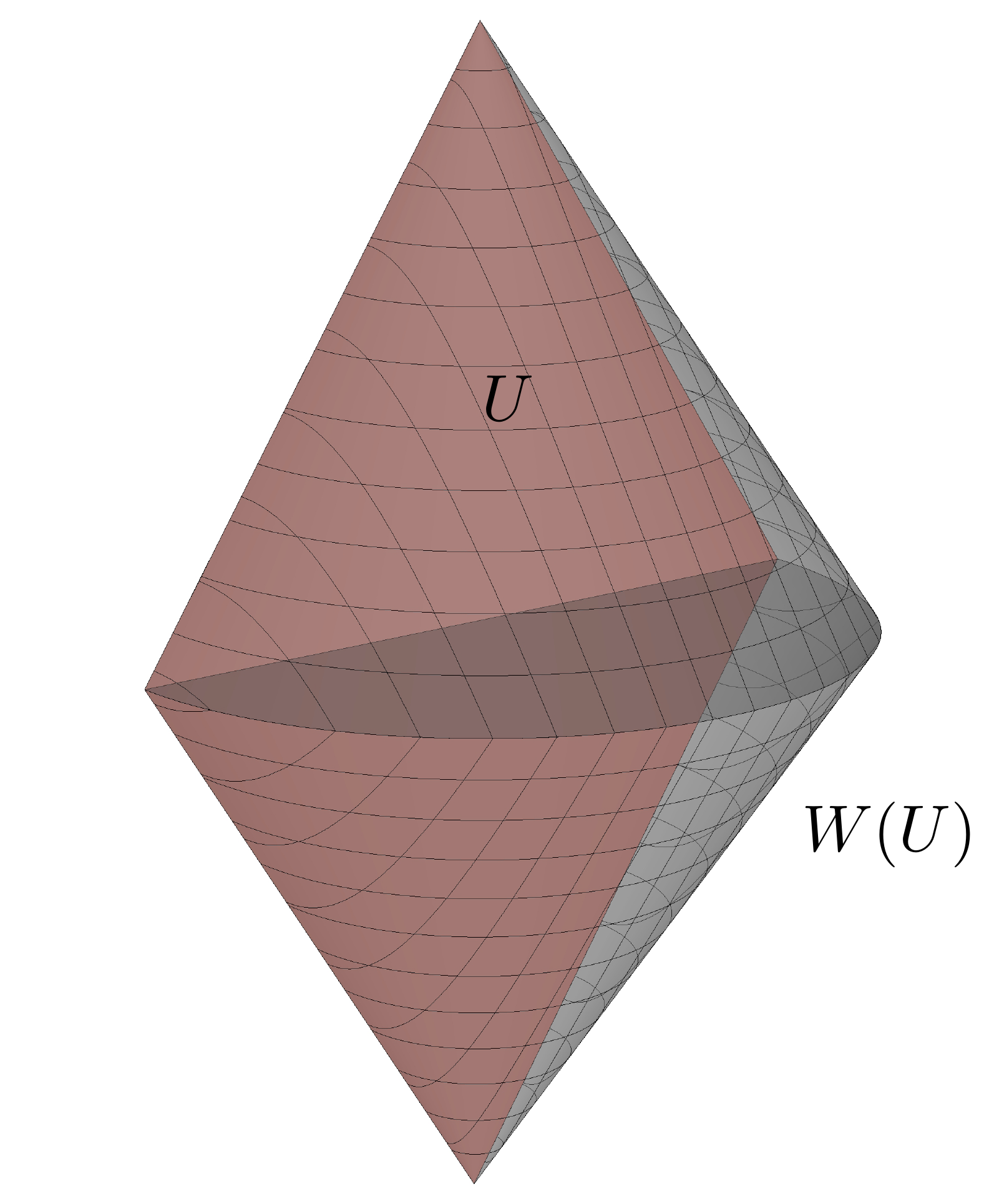} \end{center}
\caption{The boundary domain of dependece $U$ of a  boundary segment at constant time, which is a diamond-shaped region, and its corresponding causal wedge $W(U)$, which is a solid double cone.} \label{f01}
\end{figure}

Consider the Poincar\'e patch of the 3-dimensional Anti de Sitter (AdS) space and, on its boundary, a conformal field theory (CFT) in the vacuum state. In these circumstances the Ryu-Takayanagi (RT) formula \citep{Ryu:2006bv,Hubeny:2007xt},
\begin{equation}\label{RT}
S=\frac{A}{4G},
\end{equation}
is known to hold for any boundary spacelike segment. Here, $S$ is the entanglement entropy of the segment, $A$ is the length of the bulk geodesic joining the endpoints of the segment and $G$ is given in terms of the AdS radius $l$ and the CFT central charge $c$ by the holographic relation $c=3l/2G$. Suppose now that we slightly perturb the bulk geometry (maintaining the AdS asymptotics) and the CFT state. In \cite{Lashkari:2013koa,Faulkner:2013ica,VanRaamsdonk:2016exw} it was shown that, if the RT formula (\ref{RT}) continues to hold to first order for all boundary spacelike segments, 
then the metric perturbation satisfies the linearized Einstein equations. The purpose of the present section is to generalize this result as follows. Let $U$ be the boundary domain of dependence of some boundary spacelike segment, and let 
$W(U)$ denote the corresponding causal wedge, namely the intersection of the AdS causal future and past of $U$ (see figure \ref{f01}). We will show that, {\emph{if the RT formula (\ref{RT}) continues to hold to first order for all spacelike segments contained in $U$, 
then the metric perturbation satisfies the linearized Einstein equations in $W(U)$}}. This is a ``local'' generalization of the result of \cite{Lashkari:2013koa,Faulkner:2013ica,VanRaamsdonk:2016exw}, where the RT formula is not required to hold everywhere in the boundary but only in some boundary region, and the linearized Einstein equations are still recovered in a corresponding bulk region.

The proof of the above statement is just a slight modification of the argument of \cite{Faulkner:2013ica,VanRaamsdonk:2016exw}. We will thus follow essentially the same steps as in those references, although our presentation is a bit simpler (partly due to the low dimensionality) and leads to the linearized Einstein equations in a more direct and elementary way. In fact, the result we will prove is stronger than announced: {\emph{the linearized RT formula for spacelike segments contained in $U$ not only implies the linearized Einstein equations in $W(U)$, but is equivalent to those equations supplemented with a boundary condition, which relates the metric perturbation to the state perturbation via the usual holographic formula for the expectation value of the CFT stress tensor.}}

\subsection{Geometric preliminaries}

We start with a brief summary of the geometry of Poincar\'e AdS and its perturbations. In 3 dimensions, the metric of Poincar\'e AdS is
\begin{equation}\label{PAdS}
ds^2=\frac{l^2}{z^2}\left(-dt^2+dx^2+dz^2\right),
\end{equation}
with coordinate range $t,x\in{\mathbb R}$, $z>0$. This is a solution of vacuum Einstein's equations, $E_{ab}\equiv G_{ab}+\Lambda g_{ab}=0$, with cosmological constant $\Lambda=-1/l^2$. The boundary is at $z=0$, where the Minkowski metric is induced after a suitable Weyl rescaling. If $(t,x)\mapsto (t',x')$ is a boundary Poincar\'e transformation, the map $(t,x,z)\mapsto (t',x',z)$ is clearly an isometry of (\ref{PAdS}), which we also call a boundary Poincar\'e transformation. 
Together with the scale transformations $(t,x,z)\mapsto (\lambda t,\lambda x,\lambda z)$, these comprise the full group of isometries of (\ref{PAdS}). Coordinate systems in which the AdS metric takes the form (\ref{PAdS}) (thus obtained from the original one via the transformations just discussed) are called Poincar\'e coordinate systems. Any two spacelike separated boundary points $p_1$ and $p_2$ are joined by a unique bulk geodesic. In Poincar\'e coordinates in which the points are simultaneous, say $p_1=(t_0,x_0-R)$ and $p_2=(t_0,x_0+R)$, the geodesic is the semicircle
\begin{equation}\label{geo}
t=t_0\qquad (x-x_0)^2+z^2=R^2.
\end{equation}
Consider now a perturbation $\delta g_{ab}$ of (\ref{PAdS}). The perturbed spacetime is said to be asymptotically AdS if it satisfies the Brown-Henneaux boundary conditions \cite{Brown:1986nw}, which at the linearized level can be formulated as follows: there is a gauge in which $\delta g_{az}=0$ everywhere and the remaining components of $\delta g_{ab}$ are finite at the boundary.
Note that, since the boundary metric is obtained after a Weyl rescaling with Weyl factor vanishing at $z=0$, such a perturbation does not modify the boundary geometry. The first-order variation in $E_{ab}$, the left-hand side of Einstein's equations, in the above gauge is given by
\begin{alignat}{2}\label{dE}
&\delta E_{\mu\nu}=-\frac{1}{2l^2}\left[z^2\partial_{z}^2(\delta g_{\mu\nu}-\eta_{\mu\nu}\delta g)+3z\partial_z(\delta g_{\mu\nu}-\eta_{\mu\nu}\delta g)\right]\nonumber\\
&\delta E_{\mu z}=\frac{z}{l^2}\left(\partial^\nu\delta g_{\mu\nu}-\partial_\mu\delta g\right)+ O(z^2)\nonumber\\
&\delta E_{zz}=-\frac{\delta g}{l^2}+O(z),
\end{alignat}
where Greek indices correspond to the coordinates $t,x$ and are raised and lowered with the Minkowski metric $\eta_{\mu\nu}$, and $\delta g\equiv\delta g^{\mu}_{\mu}=\eta^{\mu\nu}\delta g_{\mu\nu}$. 
The omitted terms in the expansion of $\delta E_{\mu z}$ and $\delta E_{zz}$ in powers of $z$ will not be needed in the following. 

\subsection{The linearized RT formula}

Let us return to the situation described at the beginning of this section: on the boundary of Poincar\'e AdS we have a CFT in the vacuum state, and then we perturb both that state and the bulk geometry. The perturbed spacetime is required to be asymptotically AdS, so that the boundary geometry is not modified, and we work in the same gauge as above, with $\delta g_{az}=0$.
Next we will write down an explicit expression for the linearized version of the RT formula (\ref{RT}). Let $V$ be a boundary spacelike segment, and let us choose Poincar\'e coordinates in which it is at constant time, say $t=t_0$, $x\in [x_0-R,x_0+R]$. The first-order variation in the entanglement entropy of $V$ is
\begin{equation}\label{dS}
\delta S=\delta\langle H\rangle=\frac{\pi}{R}\int_{x_0-R}^{x_0+R}dx\left[R^2-(x-x_0)^2\right]\delta\langle T_{tt}\rangle(t_0,x),
\end{equation}
where $H$ is the modular Hamiltonian of $V$ in the unperturbed state (the vacuum) and $T_{\mu\nu}$ is the CFT stress-energy tensor.
The first equality above is the so-called first law of entanglement \citep{Blanco:2013joa}, which holds for any quantum system in any background state; the second equality follows from the explicit expression for the modular Hamiltonian of a segment in the vacuum state of a CFT, which was derived in \cite{Casini:2011kv}.
On the other hand, the first-order variation in the bulk geodesic distance between the endpoints of $V$ is
\begin{equation}\label{dA}
\delta A=
\frac{1}{2lR}\int_{x_0-R}^{x_0+R}dx\left[R^2-(x-x_0)^2\right]\delta g_{xx}(\gamma(x)),
\end{equation}
where $\gamma(x)=(t_0,x,\sqrt{R^2-(x-x_0)^2})$ is the AdS geodesic joining the endpoints of $V$.
This result is easily obtained from the general formula for the length of a curve after noting that, although the metric perturbation induces a small variation in the geodesic, the latter does not contribute to $\delta A$ because geodesics are extrema of the length. From the above two equations we see that the RT formula (\ref{RT}) holds to first order for $V$ if and only if
\begin{equation}\label{RT1}
\int_{x_0-R}^{x_0+R}dx\left[R^2-(x-x_0)^2\right]\left[\delta\langle T_{xx}\rangle(t_0,x)-\frac{1}{8\pi Gl}\delta g_{xx}(\gamma(x))\right]=0,
\end{equation}
where we have used that $T_{tt}=T_{xx}$ because the CFT stress tensor is traceless (recall that our metric perturbation does not modify the flat boundary geometry, so it does not give rise to a trace anomaly).
This is the linearized RT formula for a boundary spacelike segment, in Poincar\'e coordinates such that the segment is at constant time. Of course one can rewrite this formula in a coordinate-independent manner, but it will not be necessary for our purposes. If the CFT is holographic and the metric perturbation is the dual of the state perturbation, then one can check (e.g. by computing $\delta g_{\mu\nu}$ via the HKLL procedure) that the integrand in (\ref{RT1}) vanishes and hence the formula is satisfied. We emphasize, however, that we are not assuming that the CFT is holographic. We consider a generic CFT, and we look for necessary and sufficient conditions for the linearized RT formula to hold for all spacelike segments contained in a certain boundary region.


\subsection{Segments}
\label{ch:segments}

\begin{figure}[t]
\begin{center}
\includegraphics[scale=0.22]{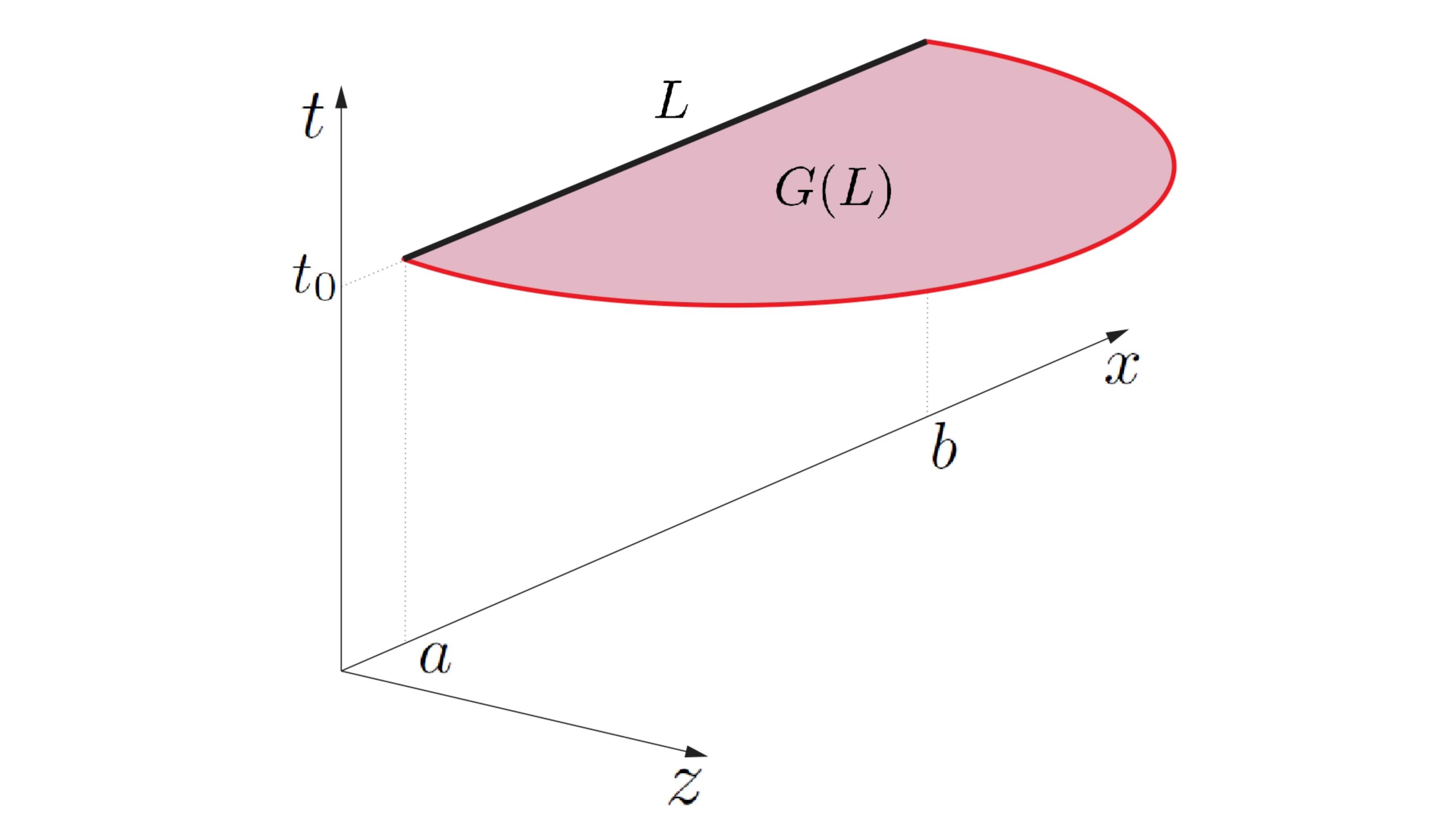} \end{center}
\caption{The union $G(L)$ of all AdS geodesics with both endpoints inside a boundary segment $L$ at constant time $t_0$. The red semicircle is the geodesic joining the endpoints of $L$.} \label{disk}
\end{figure}

Let $L$ be a boundary spacelike open{\footnote{Not including its endpoints. Unless otherwise stated, as we did here, our segments are closed.}} segment, and let us choose Poincar\'e coordinates in which it is at constant time. Next we show that the RT formula (\ref{RT}) holds to first order for all segments contained in $L$ if and only if
\begin{equation}\label{Eins1}
\delta \langle T_{xx}\rangle=\frac{1}{8\pi Gl}\delta g_{xx}{\text{ in }}L\qquad \delta E_{tt}=0{\text{ in }}G(L),
\end{equation}
where $G(L)$ denotes the union of all AdS geodesics with both endpoints in $L$. Note that, in the coordinates we are using, this is a semidisk at constant time, as shown in figure \ref{disk}. The second equation above is the $tt$ component of the linearized Einstein equations.
\\

\begin{figure}
\begin{center}
\includegraphics[scale=0.5]{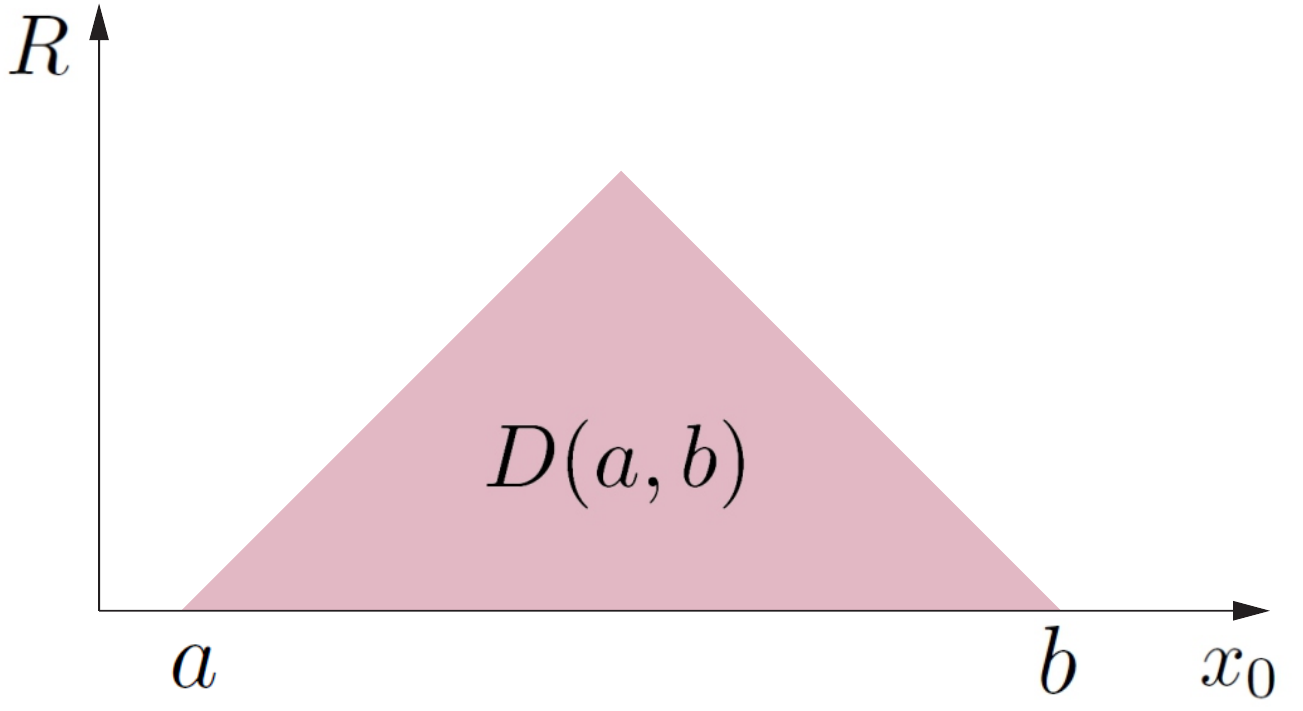} \end{center}
\caption{The set $D(a,b)$ of allowed values of the parameters $x_0$, $R$.} \label{region}
\end{figure}

Say that $L$ is the segment $t=t_0$, $x\in(a,b)$. By the results of the previous subsection, what we have to show is that the pair of local equations (\ref{Eins1}) is equivalent to the condition that the linearized RT formula (\ref{RT1}) hold for all $(x_0,R)$ in the domain
\begin{equation}
D(a,b)\equiv\{(x_0,R)\in{\mathbb R}\times{\mathbb R}^+\,|\,x_0-R>a,\,x_0+R<b\},
\end{equation}
which is represented in figure \ref{region}. Suppose first that this condition is satisfied, namely that the linearized RT formula (\ref{RT1}) holds for all
$(x_0,R)\in D(a,b)$. Then, in particular, it holds for any $x_0\in(a,b)$ provided that $R$ is sufficiently small. Evaluating this formula to leading order in the limit $R\to 0$ one easily obtains the first equation in (\ref{Eins1}), and substituting it back into (\ref{RT1}) yields
\begin{equation}\label{NL}
\int_{x_0-R}^{x_0+R}dx\left[R^2-(x-x_0)^2\right]\left[\delta g_{xx}(t_0,x,0)-\delta g_{xx}(\gamma(x))\right]=0
\end{equation}
for all $(x_0,R)\in D(a,b)$. Note that this is a purely geometric constraint. Conversely, these two equations (the first equation in (\ref{Eins1}) and (\ref{NL})) clearly imply that the linearized RT formula (\ref{RT1}) holds for all $(x_0,R)\in D(a,b)$, so they are equivalent to the latter condition.

Thus, what remains to be shown is that the non-local geometric constraint (\ref{NL}) is equivalent to the second equation in (\ref{Eins1}), namely the $tt$ component of the linearized Einstein equations in $G(L)$. 
To see this, note that the integral in (\ref{NL}) vanishes identically in the limit $R\to 0$, 
so the constraint is satisfied if and only if the derivative of that integral with respect to $R$ vanishes. 
The latter, divided by $R$, also vanishes identically in the limit $R\to 0$, so we can differentiate again to obtain the equivalent constraint
\begin{equation}\label{NL1}
\int_{x_0-R}^{x_0+R}dx\left[(\partial_{z}^2\delta g_{xx})(\gamma(x))+\frac{3}{\sqrt{R^2-(x-x_0)^2}}(\partial_z\delta g_{xx})(\gamma(x))\right]=0.
\end{equation}
The derivatives with respect to $z$ appear because there is an $R$ hidden in $\gamma$ (recall that this curve is a semicircle of radius $R$, see its expression under (\ref{dA})). Comparing with the first equation in (\ref{dE}), which gives the explicit form of $\delta E_{\mu\nu}$, we see that the above equation can be rewritten as
\begin{equation}\label{NL2}
\int_{x_0-R}^{x_0+R}dx\,\frac{1}{R^2-(x-x_0)^2}\delta E_{tt}(\gamma(x))=0.
\end{equation}
Clearly, a sufficient condition for this constraint to be satisfied for all $(x_0,R)\in D(a,b)$ is that $\delta E_{tt}=0$ in $G(L)$. To see that this condition is also necessary,
suppose that the above equation is satisfied, multiply it by $R$, integrate in this variable{\footnote{After performing these first two operations on the left-hand side of (\ref{NL2}) one obtains the integral of $\delta E_{tt}/z$ over the semidisk $t=t_0$, $(x-x_0)^2+z^2<R$.}} from $0$ to $R$ and then differentiate with respect to $x_0$. This yields an equation identical to (\ref{NL2}) except for an extra factor $x-x_0$ in the integrand, from which the term $x_0$ can be dropped using (\ref{NL2}) again. Iterating this procedure one obtains
\begin{equation}\label{NL3}
\int_{x_0-R}^{x_0+R}dx\,\frac{1}{R^2-(x-x_0)^2}x^n\delta E_{tt}(\gamma(x))=0
\end{equation}
for all $n\in{\mathbb N}$. 
Now define $f(\theta)\equiv\delta E_{tt}(\gamma(x_0+R\cos\theta))/(R\sin\theta)$ for $\theta\in (0,\pi)$. This is the function $\delta E_{tt}/z$ evaluated on the semicircle $\gamma$ and expressed in terms of the standard polar angle. The set of equations (\ref{NL3}) can then be rewritten as
\begin{equation}\label{NL4}
\int_{0}^\pi d\theta\,(\cos\theta)^nf(\theta)=0.
\end{equation}
Let now $g$ be the even extension of $f$ to the domain $(-\pi,\pi)$, i.e., $g(\theta)=f(\theta)$ for $\theta\in (0,\pi)$ and $g(\theta)=f(-\theta)$ for $\theta\in (-\pi,0)$.
Using the relation $\cos(n\theta)=T_n(\cos\theta)$, where $T_n$ denotes the Chebyshev polynomial of degree $n$, it is clear from (\ref{NL4}) that all Fourier coefficients 
of g vanish. Therefore, $g$ itself must vanish, so $f$ also vanishes and, in consequence, $\delta E_{tt}=0$ in $G(L)$ as we wanted to show.

The result just proven can be easily restated in a coordinate-independent way: the RT formula (\ref{RT}) holds to first order for all segments contained in a boundary spacelike open segment $L$ if and only if
\begin{equation}\label{Einstt+BC}
\delta\langle T_{\mu\nu}\rangle l^\mu l^\nu=\frac{1}{8\pi Gl}\delta g_{\mu\nu}\,l^\mu l^\nu{\text{ in }}L \qquad \delta E_{\mu\nu}n^\mu n^\nu=0{\text{ in }}G(L)
\end{equation}
for some boundary vector $l^\mu$ tangent to $L$ and some bulk vector $n^\mu$ normal to $G(L)$. Note that $n^z=0$ in all Poincar\'e coordinate systems (because this component is invariant under boundary boosts and $n\propto\partial_t$ when $L$ is at constant time), hence the use of Greek indices for $n^\mu$. The above equation clearly reduces to (\ref{Eins1}) in Poincar\'e coordinates such that $L$ is at constant time.

\subsection{Open regions}
\label{ch:open}

Let $U$ be a boundary open region. Next we show that the RT formula (\ref{RT}) holds to first order for all spacelike segments contained in $U$ if and only if
\begin{equation}\label{Eins+BC}
\delta\langle T_{\mu\nu}\rangle=\frac{1}{8\pi Gl}\delta g_{\mu\nu}{\text{ in }}U\qquad \delta E_{ab}=0{\text{ in }}G(U),
\end{equation}
where $G(U)$ denotes the union of all AdS geodesics joining the endpoints of some spacelike segment contained in $U$.
The first of these equations is the standard holographic formula for the expectation value of the CFT stress tensor \citep{Balasubramanian:1999re, deHaro:2000vlm}; the second is, of course, the linearized Einstein equations.

\begin{figure}
\begin{center}
\includegraphics[scale=1]{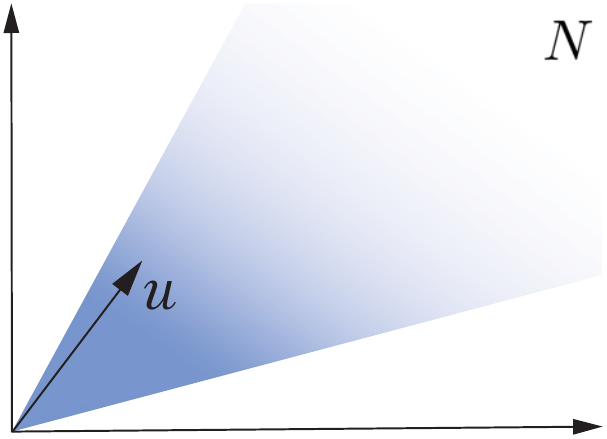} 
\end{center}
\caption{A conical neighborhood $N$ of a vector $u$ in a 2-dimensional vector space.} 
\label{vector}
\end{figure}

Since $U$ is open, any segment contained in $U$ is also contained in some open segment which is itself contained in $U$. Thus, by the results of the previous subsection, what we have to show is that (\ref{Einstt+BC}) holds for any spacelike open segment $L\subset U$ if and only if (\ref{Eins+BC}) holds.  The ``if'' part of this statement is obvious (note that $G(L)\subset G(U)$, i.e., any geodesic with both endpoints in $L$ joins the endpoints of a spacelike segment contained in $U$), so we only have to prove the ``only if'' part.
Our argument below hinges on the following simple result:
let $T$ be a symmetric rank-2 tensor over a 2-dimensional vector space $V$, and let $N\subset V$ be a conical neighborhood of some vector $u\in V$ (see figure \ref{vector}). Then,
\begin{equation}\label{conical}
T(v,v)=0{\text{ for all }}v\in N\quad\Rightarrow\quad T=0.
\end{equation}
Indeed, consider two non-collinear vectors $e_1,e_2\in N$, which form a basis of $V$. If the above hypothesis is satisfied we have $T(e_1,e_1)=T(e_2,e_2)=0$. Moreover, $e_1+e_2\in N$, so that $0=T(e_1+e_2,e_1+e_2)=2T(e_1,e_2)$ and hence $T=0$. Note that this is true regardless of how small $N$ is. This result clearly generalizes to higher dimensions; we restricted to the 2-dimensional case for simplicity and because this is the only case relevant for our purposes.

Suppose now that (\ref{Einstt+BC}) holds for any spacelike open segment $L\subset U$, and let us show that this implies (\ref{Eins+BC}). Clearly, for any point $p\in U$ and any boundary spacelike vector $l^\mu$ at $p$, there is an open segment $L\subset U$ passing through $p$ with tangent $l^\mu$, so the first equation in (\ref{Einstt+BC}) is satisfied everywhere in $U$ and for any boundary spacelike vector $l^\mu$. By (\ref{conical}), this implies the first equation in (\ref{Eins+BC}). Consider now a point $p=(t,x,z)\in G(U)$ and a spacelike open segment $L\subset U$ such that $p\in G(L)$. If $\sigma_s$ denotes the one-parameter group of boundary boosts centered at $(t,x)$, which leaves $p$ invariant, we have $p=\sigma_s(p)\in G(\sigma_s(L))$. Moreover, since $U$ is open, for sufficiently small $s$ we have $\sigma_s(L)\subset U$. Therefore, the second equation in (\ref{Einstt+BC}) is satisfied at $p$ for any $n^\mu$ in a sufficiently small conical neighborhood (within the subspace spanned by $\partial_t$ and $\partial_x$) of some vector normal to $G(L)$, so, by (\ref{conical}),
\begin{equation}\label{dEmunu}
\delta E_{\mu\nu}=0{\text{ in }}G(U).
\end{equation}
The remaining components of the linearized Einstein equations then follow from the conservation of $\delta E_{ab}$ and the first equation in (\ref{Eins+BC}), which has already been proved. Indeed, the above equation and the conservation of $\delta E_{ab}$ imply
\begin{equation}\label{cons}
\partial_z\delta \bar E_{\mu z}=0\qquad \partial_z\delta E_{zz}+z\partial^\mu\delta \bar E_{\mu z}=0 \qquad{\text{in }}G(U),
\end{equation}
where we have defined $\delta \bar E_{\mu z}\equiv \delta E_{\mu z}/z$ and, as before, Greek indices are raised and lowered with the Minkowski metric. On the other hand, since the CFT stress tensor is traceless and conserved, the first equation in (\ref{Eins+BC}) implies $\delta \bar E_{\mu z}=\delta E_{zz}=0$ in $U$ (see the explicit form of these components in (\ref{dE})). Thus we have a very simple initial value problem (with $z$ playing the role of time) for $\delta \bar E_{\mu z}$ and $\delta E_{zz}$. In order to solve it, note first that, for every point $p=(t,x,z)\in G(U)$, the point $p_0=(t,x,0)$ lies in $U$ and, furthermore, the segment joining $p_0$ and $p$ is contained in $G(U)$ (indeed, consider a spacelike segment $L\subset U$ such that $p\in G(L)$; it is clear from figure \ref{disk} that $p_0\in L$ and the segment joining $p_0$ and $p$ is contained in $G(L)$). This property enables us to integrate (\ref{cons}) in $z$ from $U$ to any point in $G(U)$, and thus conclude that the unique solution to the initial value problem is $\delta \bar E_{\mu z}=\delta E_{zz}=0$ in $G(U)$. In other words, $\delta E_{az}=0$ in $G(U)$, which completes the proof.

\subsection{Domains of dependence}
\label{ch:domains}

Let $X$ be a boundary spacelike open segment, and let $U=D(X)$ be its boundary domain of dependence, namely the set of boundary points $p$ for which any boundary inextensible causal curve containing $p$ passes through $X$. This is an open region (because $X$ is open), so the results of the previous subsection apply to it. Next we show that $G(U)=W(U)$, where $W(U)$ denotes the causal wedge of $U$ (i.e., the intersection of its AdS causal future and past), thereby completing the argument for the main statement of this section.

That $W(U)\subset G(U)$ is clear from figure \ref{f01}. Indeed, in Poincar\'e coordinates such that $X$ is at constant time, any point in $W(U)$ lies in an AdS geodesic at constant time (which is a semicircle) centered in the vertical axis of $U$, and this geodesic joins the endpoints of a segment contained in $U$. On the other hand, the inclusion $G(U)\subset W(U)$ is a consequence of entanglement wedge nesting, which was proved for generic asymptotically AdS spacetimes satisfying the null curvature condition in \cite{Wall:2012uf}. This inclusion can also be seen more directly, and very simply, as follows. Consider a spacelike segment $V\subset U$, and let $\gamma$ be the AdS geodesic joining its endpoints. From figure \ref{f01}, with $U$ replaced by $D(V)$, it is clear that $\gamma\subset W(D(V))$ (recall that, unless otherwise stated, our segments are closed, so $W(D(V))$ is closed). Moreover, from the definition of a domain of dependence it follows immediately that $D(V)\subset U$, so $W(D(V))\subset W(U)$ and hence $\gamma\subset W(U)$ as we wanted to show.

In summary, we have shown that, given a boundary spacelike open segment $X$, the RT formula (\ref{RT}) holds to first order for all spacelike segments contained in the boundary domain of dependence $U$ of $X$ if and only if
\begin{equation}\label{Einsfinal}
\delta\langle T_{\mu\nu}\rangle=\frac{1}{8\pi Gl}\delta g_{\mu\nu}{\text{ in }}U\qquad \delta E_{ab}=0{\text{ in }}W(U).
\end{equation}
Thus, the linearized RT formula for segments in $U$ implies the linearized Einstein equations in the causal wedge $W(U)$, and in fact it is equivalent to those equations supplemented with a standard holographic formula, which plays the role of a boundary condition. Some implications of this result will be discussed in section \ref{ch:discussion}.

\section{Non-zero temperature}\label{ch:btz}

The results of the previous section remain true in the case where the background bulk geometry is the planar BTZ black hole and the background CFT state is the thermal state at the black hole temperature. This is because the latter configuration is just a patch of the Poincar\'e AdS/vacuum configuration (the background considered in the previous section), as we will now explain.

\begin{figure}
\begin{center}
\includegraphics[scale=0.4]{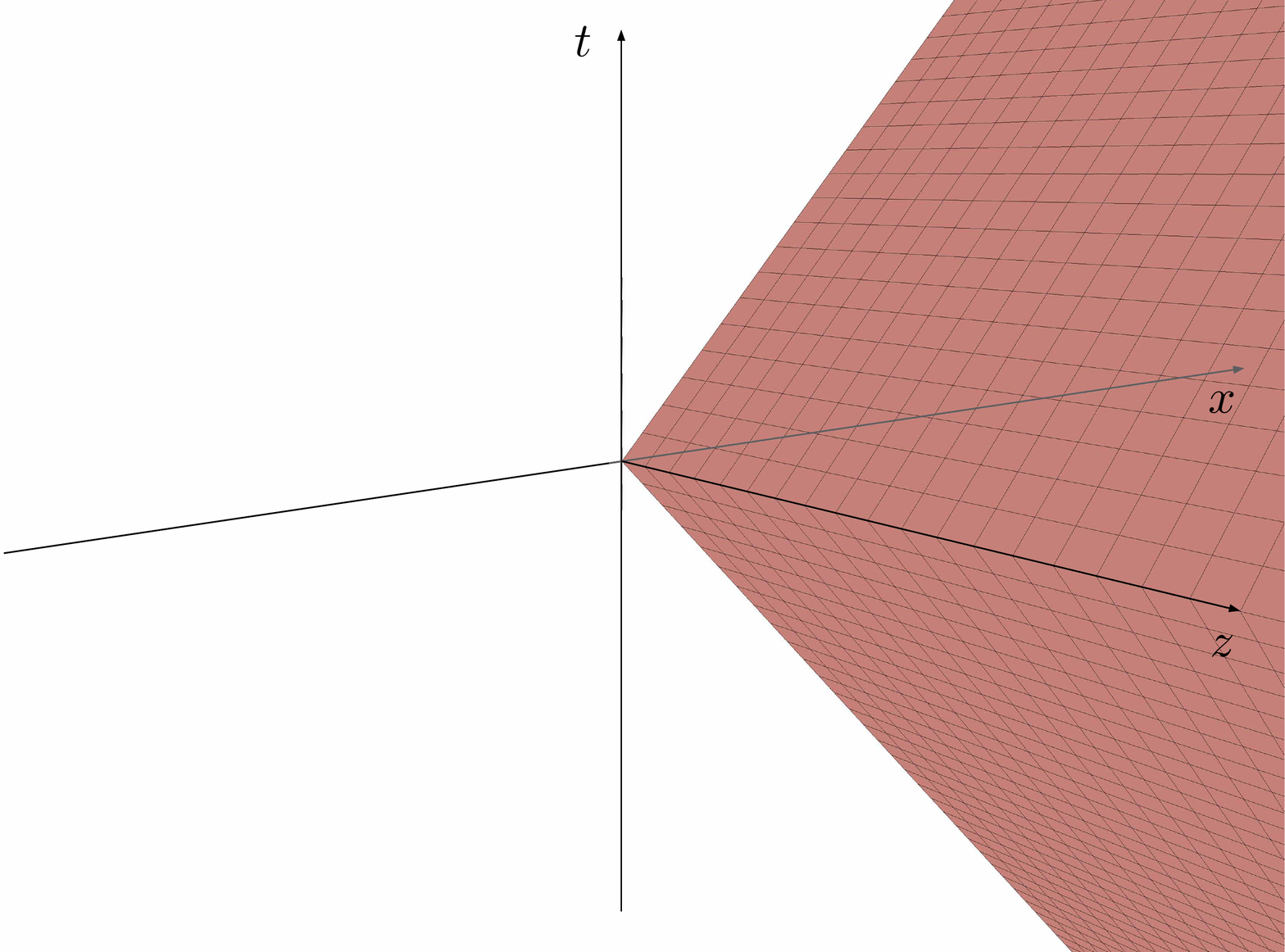} 
\end{center}
\caption{The planar BTZ black hole, viewed as a patch of Poincar\'e AdS. The boundary of this patch is the Rindler wedge of the $z=0$ surface.} 
\label{figbtz}
\end{figure}

Let us first study the relation between the planar BTZ black hole and Poincar\'e AdS. The planar BTZ black hole of inverse temperature $\beta$ is the 3-dimensional spacetime with metric
\begin{equation}\label{btz}
ds^2=-\left(\frac{\bar r^2}{l^2}-M\right)d\bar t^2+\frac{d\bar r^2}{\frac{\bar r^2}{l^2}-M}+\frac{\bar r^2}{l^2}d\bar x^2,
\end{equation}
where $M=(2\pi l/\beta)^2$ and the coordinate range is $\bar t,\bar x\in{\mathbb R}$, $\bar r>l\sqrt M$ (note that what we are calling the planar BTZ black hole is, more precisely, the region of the planar BTZ black hole outside the horizon). This is just a patch of Poincar\'e AdS. Indeed, setting
\begin{alignat}{2}\label{coord}
&t=l\sqrt{1-\frac{M l^2}{\bar r^2}}e^{\sqrt M\bar x/l}\sinh(\sqrt M\bar t/l)\nonumber\\
&x=l\sqrt{1-\frac{M l^2}{\bar r^2}}e^{\sqrt M\bar x/l}\cosh(\sqrt M \bar t/l)\nonumber\\
&z=\frac{\sqrt M l^2}{\bar r}e^{\sqrt M\bar x/l}
\end{alignat}
in the Poincar\'e AdS metric (\ref{PAdS}) one recovers the planar BTZ metric (\ref{btz}). The domain of this coordinate transformation is $\bar t,\bar x\in{\mathbb R}$, $\bar r>l\sqrt M$, and its image is $|t|<x$, $z>0$. Therefore, the planar BTZ black hole is the region $|t|<x$ of Poincar\'e AdS, see figure \ref{figbtz}. In particular, the boundary of the planar BTZ black hole is the Rindler wedge of the boundary of Poincar\'e AdS.

On the boundary of the planar BTZ black hole, consider a CFT at the black hole temperature. The corresponding modular Hamiltonian $H$, which is defined in terms of the density matrix $\rho$ by the equation $\rho=e^{-H}/\Tr(e^{-H})$, is thus $\beta$ times the true Hamiltonian of the theory. In other words, $H$ is the Noether charge associated with the Killing vector $\xi=\beta\partial_{\bar t}$. In terms of Poincar\'e coordinates,
\begin{equation}
\xi=\beta\left(\frac{\partial t}{\partial \bar t}\,\partial_{t}+\frac{\partial x}{\partial \bar t}\,\partial_{x}\right)=2\pi\left(x\partial_{t}+t\partial_{x}\right),
\end{equation}
so we may alternatively say that the CFT is at inverse temperature $2\pi$ with respect to the standard boost generator on the Rindler wedge, which is the vector between parentheses on the right-hand side above. Now, let us view the theory as being defined on the whole boundary of Poincar\'e AdS and the thermal state as a reduced density matrix obtained by tracing out the degrees of freedom outside the Rindler wedge. By the Bisognano-Wichmann theorem \cite{Bisognano:1976za}, we may take the global state of the theory to be the vacuum state.

Thus we have seen that the planar BTZ black hole/thermal state configuration is the patch $|t|<x$ of the Poincar\'e AdS/vacuum configuration, so the results of the previous section are also valid for perturbations of this thermal background. In particular, since the Rindler wedge is the domain of dependence of a segment with one of its endpoints sent to infinity and its corresponding causal wedge is the entire planar BTZ black hole (compare figures \ref{f01} and \ref{figbtz}), if the linearized RT formula holds for all boundary spacelike segments{\footnote{We are considering perturbations of the thermal background, so by boundary we mean that of the planar BTZ black hole.}} then the metric perturbation satisfies the linearized Einstein equations throughout the bulk. Note that we would not have been able to reach this conclusion directly from the results of \cite{Lashkari:2013koa,Faulkner:2013ica,VanRaamsdonk:2016exw}, because there the linearized RT formula was assumed to hold for all spacelike segments in the entire boundary of Poincar\'e AdS.

\section{Discussion}
\label{ch:discussion}

In this paper we have considered the 3-dimensional Poincar\'e AdS space and, on its boundary, a CFT in the vacuum state, a situation in which, by explicit computation, the RT formula (\ref{RT}) is known to hold for all boundary spacelike segments. Then we have perturbed both the bulk geometry and the CFT state, and we have studied under what conditions the RT formula continues to hold to first order for all spacelike segments contained in a boundary region $U$, which we have taken to be the domain of dependence of some given spacelike segment. We have found that a necessary and sufficient condition is that the metric perturbation satisfy (i) the linearized Einstein equations in the causal wedge $W(U)$, and (ii) a boundary condition which relates its value at $U$ to the state perturbation via the usual holographic formula for the expectation value of the CFT stress tensor (see (\ref{Einsfinal}) for the explicit equations and figure \ref{f01} for a representation of the regions $U$ and $W(U)$). We have also shown that the same is true for small perturbations of the planar BTZ black hole and the CFT thermal state.

These results generalize the analysis of \cite{Lashkari:2013koa,Faulkner:2013ica,VanRaamsdonk:2016exw}, where the linearized Einstein equations where first shown to follow from the linearized RT formula, by weakening its assumptions: the latter formula is not required to hold everywhere in the boundary but only in some boundary region, and the background state is allowed to have any temperature. Our argument is similar to that of \cite{Faulkner:2013ica,VanRaamsdonk:2016exw}, although it is perhaps a bit simpler in some steps, and we have stressed that it shows not only the implication ``linearized RT $\Rightarrow$ linearized Einstein + boundary condition'' but also the converse one.
The analysis of \cite{Lashkari:2013koa,Faulkner:2013ica,VanRaamsdonk:2016exw}, unlike ours, is not restricted to the case of 3 bulk dimensions, but is valid for any bulk dimensionality{\footnote{In some sense, the derivation of the linearized Einstein equations from the linearized RT formula is a stronger result in $D\ge 5$ bulk dimensions than in lower dimensions, because, for $D<5$, the Lovelock theorem states that the only diffeomorphism-invariant theory of gravity which has second-order equations of motion is Einstein gravity.}}. There seems to be no obstruction for our zero-temperature results to carry over to higher dimensions, but this is not so clear in the case of non-zero temperature, where our arguments relied heavily on the bulk being 3-dimensional.

It is worth noting that the above boundary value problem, Eq.~(\ref{Einsfinal}), has a unique solution. Indeed, from the first equation in (\ref{dE}) one easily sees that the $\mu\nu$ components of the linearized Einstein equations and the requirement that $\delta g_{\mu\nu}$ be finite at the boundary constrain the metric perturbation in $W(U)$ to be independent of $z${\footnote{This is true both at zero and non-zero temperature. Note from (\ref{coord}) that, in the case of non-zero temperature, the metric perturbation depends on the BTZ radial coordinate $\bar r$ even though it does not depend on $z$.}}, and thus to be completely determined by the boundary condition (the remaining components of the linearized Einstein equations are just boundary value constraints, which are satisfied by our specific boundary condition). Therefore, {\emph{given a perturbation of the CFT state, requiring that the linearized RT formula be satisfied for all spacelike segments contained in $U$ determines completely the metric perturbation in $W(U)$}}. This result is a very explicit and simple example of bulk geometry emerging from boundary entanglement. It also gives further evidence for subregion-subregion duality, namely the idea that, in holography, a boundary domain of dependece contains complete information about some corresponding bulk region \cite{Bousso:2012sj, Bousso:2012mh, Czech:2012bh}. The latter is believed to be the so-called entanglement wedge rather than the causal wedge (see \cite{Czech:2012bh,Headrick:2014cta,Dong:2016eik} for some evidence in this direction), but both bulk regions coincide in the case we have been considering, where the boundary region is the domain of dependence of a segment.

Our analysis, however, also applies to boundary regions $U$ for which the causal wedge and the entanglement wedge do not coincide. Suppose, for example, that we require the linearized RT formula to hold for all spacelike segments contained in a boundary region $U$ which is the union of two non-overlapping regions, each of which is the domain of dependence of a spacelike segment. 
According to what has been seen above, this determines completely the metric perturbation in the union of the corresponding causal wedges. If the segments are spacelike separated, then $U$ is itself a domain of dependence, and the union of causal wedges, where the metric perturbation is determined, is the causal wedge of $U$. If the segments are sufficiently close to each other, the entanglement wedge is larger (always containing the causal wedge), so there is a part of this bulk region where the geometry is not reconstructed. This is not in contradiction with subregion-subregion duality; it simply says that the RT formula for segments alone is not enough to reconstruct the geometry everywhere in the entanglement wedge but only in the causal wedge. It is plausible that the geometry may be reconstructed past the causal wedge and throughout the entanglement wedge by imposing a stronger condition, for example that the RT formula hold not only for segments but also for unions of segments. We leave the study of this possibility for future work{\footnote{This is a difficult problem, because the vacuum modular Hamiltonian of a union of segments is not known for generic CFTs, and in the cases where it is known it includes non-local contributions \cite{Casini:2009vk}. The problem might be more tractable in the limit of large central charge, which, after all, is the relevant limit for holography. For example, in this limit the vacuum entanglement entropy can be computed for arbitrary unions of intervals \cite{Hartman:2013mia}.}}.

Let us now comment on the implication ``linearized Einstein + boundary condition $\Rightarrow$ linearized RT'', which, as emphasized above, is part of what we have shown in this paper. For CFTs with a gravity dual, a perturbation of the vacuum or a thermal state has associated a bulk metric perturbation which certainly satisfies the linearized Einstein equations and the boundary condition (which is a standard holographic formula). Therefore, the above implication tells us that the linearized RT formula for segments is satisfied in holographic CFTs for perturbations of the vacuum or  a thermal state. This is, of course, not new: the RT formula is already known to hold in holographic CFTs, for generic states and boundary spatial regions \cite{Lewkowycz:2013nqa, Dong:2016hjy}. We just point out that our arguments serve as an alternative holographic proof of the RT formula, at the linearized level and for boundary segments.

As for future prospects, we have already mentioned one: study how the metric perturbation is constrained when the linearized RT formula is imposed not only for segments but also for unions of segments, and see if this allows to reconstruct the geometry past the causal wedge and throughout the entanglement wedge in the cases where these two bulk regions do not coincide. Another future direction is to extend the analysis of this paper to other measures of entanglement for which holographic recipes have been proposed, of which a recent example is the entanglement of purification \cite{Takayanagi:2017knl, Nguyen:2017yqw, Bao:2017nhh}.

\section*{Acknowledgements}

It is a pleasure to thank Joan Camps, Horacio Casini, Eoin Colgain, Gaston Giribet, Alex May, Felipe Rosso and Rapha\"el Turrents for useful discussions. This work was partially supported by CONICET, Universidad de Buenos Aires, Instituto de Astronom\'ia y F\'isica del Espacio and Fundaci\'on Bunge y Born. The authors are also grateful to the organizers of the Bariloche ``It from Qubit'' workshop, where part of this work was developed.

\bibliography{bibbtz}
\bibliographystyle{JHEP}
\end{document}